\documentclass[aps,prb,reprint,superscriptaddress]{revtex4-2}

\usepackage{graphicx} 
\usepackage{color} 
\usepackage{hyperref} 
\usepackage{dcolumn} 
\usepackage{soul} 

\begin{document}

\title{Effective decoupling of ferromagnetic sublattices by frustration in Heusler alloys}

\author{F. Cugini}\email[corresponding author:]{francesco.cugini@unipr.it}
\affiliation{Department of Mathematical, Physical and Computer Sciences, University of Parma, Parco Area delle Scienze 7/A, 43124 Parma, Italy} 
\affiliation{Institute of Materials for Electronics and Magnetism, National Research Council (IMEM-CNR), Parco Area delle Scienze 37/A, 43124 Parma, Italy}

\author{S. Chicco}
\affiliation{Department of Mathematical, Physical and Computer Sciences, University of Parma, Parco Area delle Scienze 7/A, 43124 Parma, Italy}
\author{F. Orlandi}
\affiliation{ISIS Pulsed Neutron Facility, STFC, Rutherford Appleton Laboratory, Chilton, Didcot, Oxfordshire OX11-0QX, United Kingdom}
\author{G. Allodi}
\author{P. Bonf\'a}
\author{V. Vezzoni}
\affiliation{Department of Mathematical, Physical and Computer Sciences, University of Parma, Parco Area delle Scienze 7/A, 43124 Parma, Italy} 
\author{O. N. Miroshkina}
\author{M. E. Gruner}
\affiliation{Department of Physics and Center for Nanointegration, CENIDE, University of Duisburg-Essen, Lotharstr. 1, 47048 Duisburg, Germany}
\author{L. Righi}
\affiliation{Department of Chemistry, Life Sciences and Environmental Sustainability, University of Parma, Parco Area delle Scienze 11/A, 43124 Parma, Italy} 
\affiliation{Institute of Materials for Electronics and Magnetism, National Research Council (IMEM-CNR), Parco Area delle Scienze 37/A, 43124 Parma, Italy}
\author{S. Fabbrici}
\author{F. Albertini}
\affiliation{Institute of Materials for Electronics and Magnetism, National Research Council (IMEM-CNR), Parco Area delle Scienze 37/A, 43124 Parma, Italy}
\author{R. De Renzi}
\affiliation{Department of Mathematical, Physical and Computer Sciences, University of Parma, Parco Area delle Scienze 7/A, 43124 Parma, Italy} 
\author{M. Solzi}
\affiliation{Department of Mathematical, Physical and Computer Sciences, University of Parma, Parco Area delle Scienze 7/A, 43124 Parma, Italy} 
\affiliation{Institute of Materials for Electronics and Magnetism, National Research Council (IMEM-CNR), Parco Area delle Scienze 37/A, 43124 Parma, Italy}

\date{\today}

\begin{abstract}
Magnetic frustration in ferromagnetic metallic systems is unusual due to the long-range and symmetric nature of the exchange interactions. In this work we prove that it is possible to obtain a highly frustrated ferromagnetic phase in a multi-sublattices cubic structure through a fine tuning of the magnetic interactions. This peculiar state is achieved in Ni-Mn-(In, Sn) Heusler alloys and results in the effective decoupling of their two intertwined ferromagnetic sublattices. One sublattice is ferromagnetic long range ordered below the macroscopic Curie temperature ($T_C$) whereas the second one remains disordered until a crossover to a polarized state occurs at $T<<T_C$. This result points out that a fine engineering of the magnetic interactions in metallic systems can lead to interesting novel and emergent phenomena.
\end{abstract}

\maketitle

Frustration in magnetic systems is the consequence of particular combinations of lattice geometry and competing magnetic interactions, and it is known to drive interesting exotic and collective phenomena.\cite{gilbert} The simplest frustrated model is described by antiferromagnetic interactions on a triangular lattice, where not all the bonds can be satisfied simultaneously causing a degenerate ground state. Over the years more complex models and materials with frustrated configurations were discovered and studied.\cite{Bramwell2001,Taguchi2001,Kosuke2018,Castelnovo2008} Most of them arise from localized moments in insulating antiferromagnetic materials, whereas frustration in ferromagnetic metallic systems is less common.\cite{Stockert2020} The situation becomes even more interesting in the case of intertwined magnetic sublattices, where the balance and the possible frustration of inter- and intra-sublattices exchange interactions can induce the decoupling of the magnetic ordering.\cite{Wills,Damay,Garlea,Cadogan,Morrow,Lyubutin,Orlandi} When the intra-sublattices interaction prevails, the magnetic sublattices behave independently and might order at different temperatures with different periodicity.\cite{Wills} This scenario rarely occurs in ferromagnetic materials, since the periodicity of the magnetic structure coincides with the nuclear one and the effective magnetic field that one sublattice exerts on the others does not generally vanish.

In this work we show that it is possible to obtain a highly frustrated state in a two sublattices ferromagnetic metal.  We design this peculiar state by finely controlling the magnetic interactions of a cubic off-stoichiometric Ni-Mn-(In,Sn) alloy through electron doping of the non-spin-polarized atom. The frustrated magnetic configuration exhibits one sublattice showing ferromagnetic long range ordering below the Curie temperature ($T_C$) of the material whereas the second sublattice remains disordered until a crossover to a polarized state occurs at $T<<T_C$.

\begin{figure*}
	\centering
	\includegraphics[width=1\textwidth,keepaspectratio]{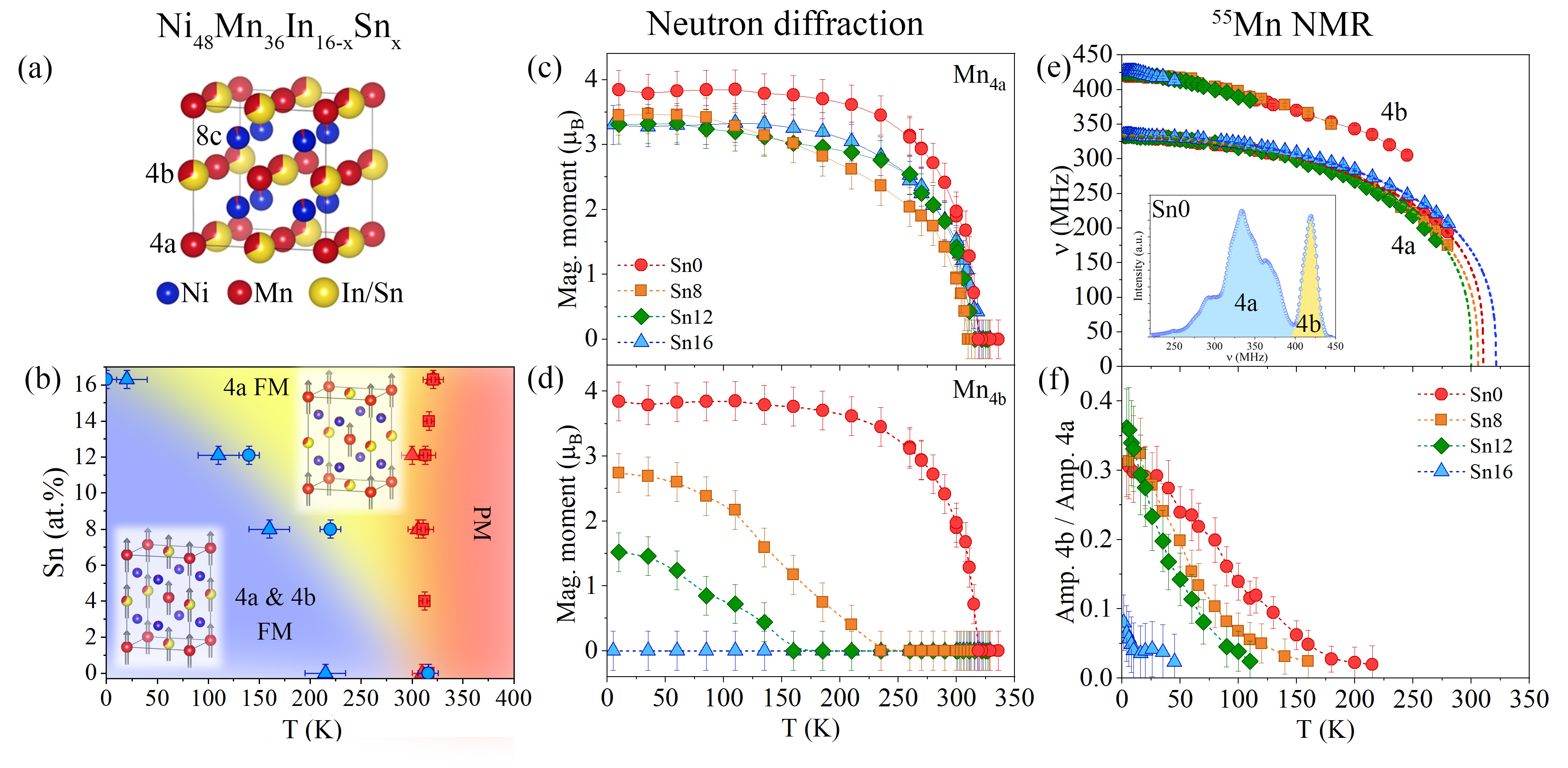}
\caption{\label{1} (a) Crystal structure of the paramagnetic austenite phase of the Ni$_{48}$Mn$_{36}$In$_{16-x}$Sn$_x$ alloys (b) Ni$_{48}$Mn$_{36}$In$_{16-x}$Sn$_x$ magnetic phase diagram. The critical temperatures of 4a and 4b sublattices are obtained from magnetometry (squares), refinement of NPD (circles) and NMR data (triangles). The NMR critical temperature of the 4a sublattice is derived from the best fit of the $\nu(T)$ data to the empirical law $\nu(T)=\nu_{0}[1-(T/T_{c})^{\alpha}]^{\beta}$ (panel e, dashed curves), whereas the onset temperature of the 4b sublattice is defined as the temperature at which the 4b peak disappears. (c-d) Mn magnetic moment in the (c) 4a  and (d) 4b sublattices as a function of temperature, obtained from the refinement of NPD data. (e) Zero-field 4a and 4b mean $^{55}$Mn resonance frequencies vs temperature. Inset: $^{55}$Mn resonance spectrum of Sn0 at $T = 10$ K. (f) Ratio of integrated 4b and 4a NMR peak amplitudes as a function of temperature.}
\end{figure*}

Ni$_{50}$Mn$_{25+x}$Z$_{25-x}$ (Z= In, Sn, Ga, ...) Heusler alloys are intensively investigated thanks to the strong coupling between their structural, magnetic and electronic degrees of freedom, which gives rise to a great variety of functional properties.\cite{ullakko,liu,manosa,graf,zhang,yu,Fabbrici2022} These compounds crystallize in a cubic \textit{${L2_1}$} austenite structure (Fig. \ref{1}a), composed of four intertwined face-centered cubic sublattices, with a $T_C$ in the 300-400 K range. The ferromagnetic ground state is due to the superposition of short-range and long-range magnetic interactions between Mn and Ni moments, with the Mn atoms providing the largest contribution to the magnetization, carrying a moment of about 4~$\mu_B$. \textit{Ab-initio} calculations \cite{sasioglu2004} indicate that the dominant coupling is the ferromagnetic (FM) nearest-neighbors exchange between Mn and Ni atoms in the 4a and 8c Wyckoff positions of the cubic cell (Fig. \ref{1}a).\cite{Note} Direct exchange interactions between Mn atoms on the same sublattice are irrelevant, owing to the large Mn--Mn distance ($>$ 4 \AA). Conversely, for Mn-rich alloys ($x>0$) a direct antiferromagnetic exchange can be established between the excess of Mn atoms, sited at the 4b position, and Mn atoms in the 4a sublattice.\cite{buchelnikov} Moreover, long-range super-exchange interactions, with a Ruderman-Kittel-Kasuya-Yosida-like oscillatory character, are promoted by the hybridization of \textit{d}-Mn and \textit{sp}-Z electronic states and are mediated by conduction electrons. \cite{sasioglu2008,buchelnikov}

\begin{table*}
 \caption{\label{t1} Sample composition obtained from energy dispersive x-ray spectroscopy (EDX). Cubic cell parameter at room temperature $a$. Curie temperature $T_c$, total magnetic moment per f.u. $\mu_s$ at $T = 2$~K and $\mu_0H = 5$~T obtained by magnetometry. Magnetic moment per Mn/Ni atom on 4a, 4b and 8c site derived from NPD at $T = 10$~K. Magnetic susceptibility at high field $\chi_{HF}$ at $T = 2$~K. \label{Tab1}}
 \begin{ruledtabular}
 \begin{tabular}{c c c c c c c c c}
Sample &	Composition &	$a$ &		$T_c$ &	$\mu_s$ &	$\mu_{4a}$ 	&	$\mu_{4b}$ &	$\mu_{8c}$ &	$\chi_{HF}$ \\
&	EDX ($\pm 0.5$ at. \%) &	(Å) &	$(K)$ &	$(\mu_B/f.u.)$ &	$(\mu_B/Mn)$ &	$(\mu_B/Mn)$ &	$(\mu_B/Ni)$ &	$(10^{-6} m^3kg^{-1})$	\\
\hline
Sn0  & Ni$_{47.6}$Mn$_{36.4}$In$_{16}$              & 6.02597(4)   &  316.5(5) &  5.94(6) & 3.84(4) & 3.84(4) & 0.29(1) & 0.09(2)\\
Sn4  & Ni$_{47.6}$Mn$_{36.6}$In$_{11.8}$Sn$_{4}$    &	-	           &  312.5(5) &	5.99(6) &	-       &	-       &	-       &	0.09(2)\\
Sn8  & Ni$_{47.8}$Mn$_{36.4}$In$_{7.8}$Sn$_{8}$     &	6.01701(5)	 &  310.9(5) &	5.95(6) &	3.45(3) &	2.74(7) &	0.28(1) &	0.12(3)\\
Sn12 & Ni$_{47.3}$Mn$_{36.6}$In$_{4}$Sn$_{12.1}$    & 6.01297(4)   &	313.3(5) &	5.68(6) &	3.31(4) &	1.52(8) &	0.27(1) &	0.42(1)\\
Sn14 & Ni$_{47.5}$Mn$_{36.4}$In$_{1.9}$Sn$_{14.2}$  &	6.01023(4)   &	316.7(5) &	5.21(5) &	-       & 	-     & -       &	0.69(3)\\
Sn16 & Ni$_{47.5}$Mn$_{36.2}$Sn$_{16.3}$	          &	6.00803(8)   &	320.9(5) &	4.92(5)	& 3.30(1) &	0       &	0.31(2) &	1.04(7)\\
\end{tabular}
 \end{ruledtabular}
\end{table*}

Polycrystalline samples of Ni$_{48}$Mn$_{36}$In$_{16-x}$Sn$_x$ ($0 \le x \le 16$, hereafter: Sn0-Sn16)  were prepared following standard procedure (reported in the Supplementary Material \cite{SUP,Cavazzini,Cugini}). X-ray diffraction patterns, collected at room temperature, confirmed the \textit{${L2_1}$} cubic structure and indicate a small monotonic decrease of the lattice parameters with the Sn content (Tab. \ref{t1}). \cite{Cavazzini} Magnetometry measurements revealed a non-monotonic trend of $T_C$ with the In/Sn ratio, which shows a minimum at 310.9 $\pm$ 0.5 K for Sn8 and reaches the maximum 320.9 $\pm$ 0.5 K at the end-member composition Sn16 (Tab. \ref{t1}, Fig. S3b \cite{SUP}). On the contrary, the total magnetic moment (at $T = 2$ K and $\mu_0H = 5$ T) is constant until  $x = 8$ ($\mu \sim 6.0 \mu_B/f.u$) and then drops upon rising the Sn content ($\mu = 4.92 \pm 0.05 \mu_B/f.u$ for Sn16) (Tab. \ref{t1}, Fig. S3b \cite{SUP}).  In addition, the high-field susceptibility, measured at 2 K, linearly increases by decreasing the magnetization (Tab. \ref{t1}, Fig. S4 \cite{SUP}).

Neutron powder diffraction data, collected for four samples (Sn0, Sn8, Sn12, Sn16) on the WISH diffractometer at the ISIS facility UK,\cite{Chapon2011} confirmed that the excess Mn atoms occupy the In/Sn 4b site ($\approx 36\%$) and the Ni 8c site ($\approx  4\%$), whereas the 4a site is fully occupied by Mn (Fig. \ref{1}a). Below $T_c$ the $111$ reflection intensity increases with decreasing temperature (Fig. S5 \cite{SUP}) without the appearance of new reflections, indicating the presence of long range magnetic order with $k = 0$ propagation vector. The refinement of the data, with the tetragonal magnetic space group \textit{I}4\textit{/mm'm'}, \cite{Orlandi2,Petricek2014,Campbell2006} revealed the presence of ordered magnetic moments on both Mn$_{4a}$ ($\mu \sim 3.3-3.8 \mu_B$) and Ni$_{8c}$ ($\mu \sim 0.3 \mu_B$) sublattices, for all compositions. The Mn$_{\mathrm{4a}}$ moment (Fig. \ref{1}c) increases with decreasing temperature, following a Brillouin-like behavior, comparable to the high field $M(T)$ measurements (Fig. S2b \cite{SUP}). Instead, the Mn$_{\mathrm{4b}}$ magnetic sublattice shows the same temperature dependence and moment size of the Mn$_{\mathrm{4a}}$ only for the In end-member (Sn0) (Fig. \ref{1}d). By increasing the Sn content, the onset temperature for long-range magnetic order of the 4b sublattice drops from the macroscopic Curie temperature, falling below the minimum measured temperature ($T = 10$ K) in the case of the Sn-ternary alloy (Sn16). Instead, in the latter sample, an increase of the low-Q signal for $T < 150$ K (Fig. S11 \cite{SUP}) suggests the appearance of short-range FM correlations.

$^{55}$Mn broadband nuclear magnetic resonance spectra as a function of temperature clarify and bolster the neutron diffraction results (details are reported in the Supplementary Material \cite{SUP}, see also references \cite{Allodi2005,Hahn1950,Riedi1989,Allodi2014,Freeman,Wurmehl2008,Schaf1983,Perdew1996,Methfessel1989,Okhotnikov2016,Lancaster2014,Allodi2001,Zhao2000}). The low-temperature zero-field (ZF) spectra exhibit two main distinct peaks (Fig. \ref{1}e) in the 250-450 MHz frequency interval, which are unambiguously assigned to Mn nuclei at the 4a and 4b sites. \cite{SUP} The fine structure of the majority peak (4a) reflects the disordered magnetic surrounding of $Mn_{4a}$ nuclei due to the random occupation of the 4b site by Mn.\cite{SUP} Both peaks rigidly shift to lower frequencies by applying an external magnetic field, demonstrating that both are due to ferromagnetically ordered Mn ions with moment collinear to the external magnetic field (Fig. S13 \cite{SUP}). The mean spontaneous frequency, proportional to the mean local moment, and the integrated amplitude of the 4a peak follow, for all the samples, the typical temperature behavior of ferromagnetic materials, both vanishing at the Curie temperature (Fig. \ref{1}e). Instead, the intensity of the 4b minority peak drops on warming well below the Curie temperature (Fig. \ref{1}f), while its mean frequency does not vanish (Fig. \ref{1}e). This indicates a gradual decrease of the population of magnetic ordered ions in the 4b sublattice rather than a decrease of the localized moment. This effect shifts to lower temperature by increasing the Sn content (Fig. \ref{1}f). The Sn16 NMR spectrum at 1.4 K reveals that only the 25\% of Mn moments at the 4b site (9\% of the total 4b sites) is ordered along the net moment direction and this fraction rapidly decreases by increasing the temperature.

\begin{figure}
	\centering
	\includegraphics{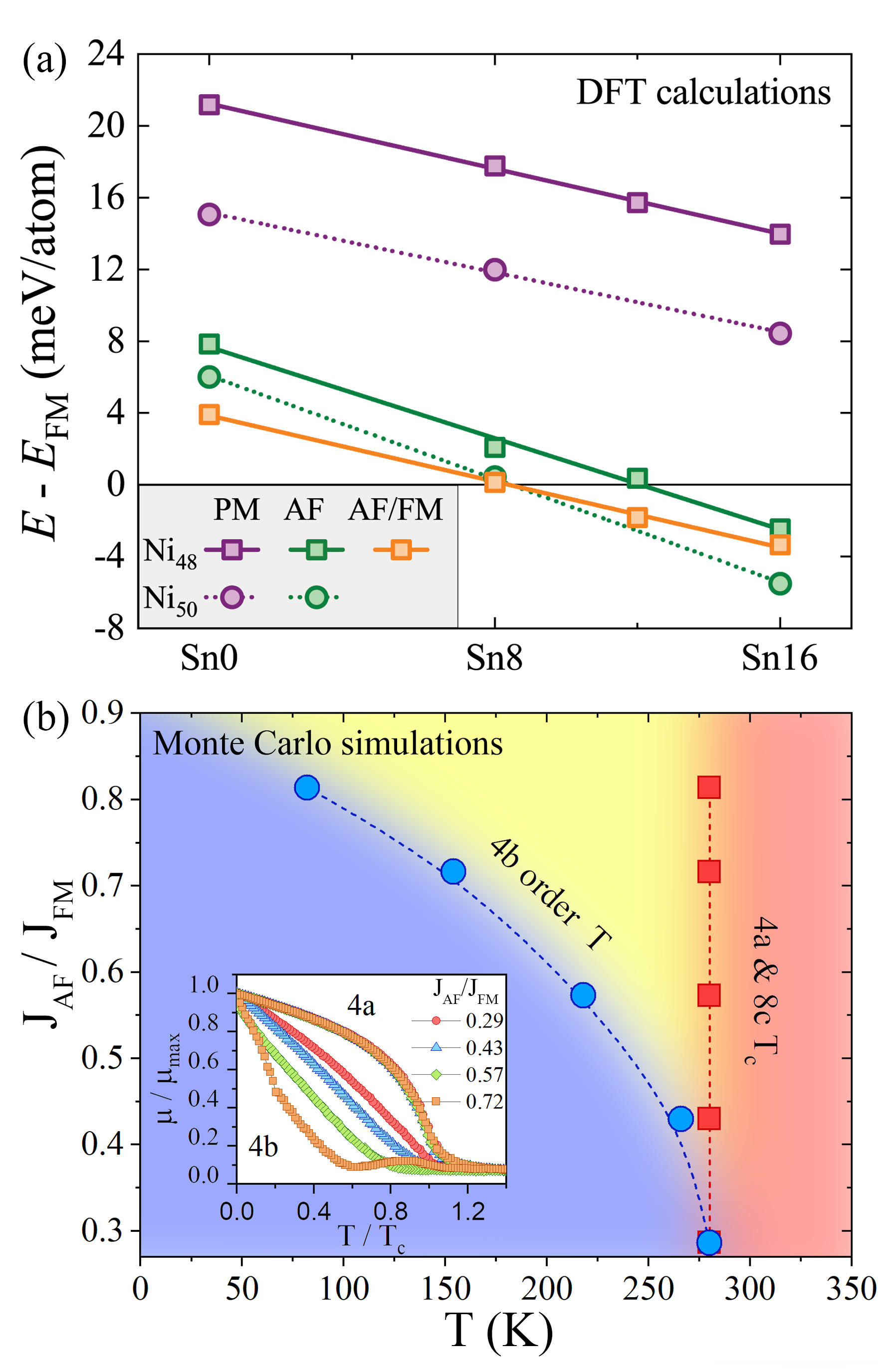}
\caption{\label{2} (a) Total energies of different magnetic arrangements (AF, AF/FM, PM) relative to the FM state of the respective composition (black line) as a function of the Sn content from first-principles DFT calculations. In Ni$_{50}$Mn$_{34}$In$_{16-x}$Sn$_x$ a transition from FM ground state to an AF coupled 4b sublattice (green circles) is predicted for $x = 8$, while Mn-excess in Ni$_{48}$Mn$_{36}$In$_{16-x}$Sn$_x$ rather leads to a mixed AF/FM arrangement on the 4b sites (orange squares) towards the Sn-rich side, favored over the AF arrangement (green squares).  Disordered magnetic arrangements (PM, purple) are significantly higher in energy. (b) Magnetic phase diagram obtained from Monte Carlo simulations by varying the $J_{AF}/J_{FM}$ ratio. Inset: temperature dependence of the normalized magnetization in the 4a and 4b sublattices. }
\end{figure}

First-principles calculations in the framework of DFT reveal the mechanism behind the effective magnetic decoupling between the 4a and 4b sublattices (details are reported in the Supplementary Material  \cite{SUP}, see also Refs. \cite{Kresse1996,Kresse1999,Blochl1994,Monkhorst1976,Schleicher2017,Liechtenstein1987,Ebert2011}). The hybridization with \textit{sp}-electrons obtained by tailoring the In/Sn content determines the position of characteristic features in the $d$-metal density of states. This has in turn consequences for the direct exchange between nearest (Ni--Mn) and next-nearest (Mn$_{\mathrm{4a}}$--Mn$_{\mathrm{4b}}$) neighbors. In this way, the effective coupling of the Mn$_{\mathrm{4b}}$ moments to the Mn$_{\mathrm{4a}}$ and Ni sublattices is progressively reduced by replacing In with Sn due to the strengthening of the AF Mn$_{\mathrm{4a}}$--Mn$_{\mathrm{4b}}$ coupling and the weakening of the competing FM exchange between Mn (4a and 4b) and Ni atoms. This effect is partially compensated by the small fraction (4\%) of Mn atoms sited on the 8c Ni sites, which show a strong AF coupling with nearest neighbors Mn$_{\mathrm{4a,4b}}$ mediating an effective FM interaction between them. These competing interactions and their change with the In/Sn content explain the different ordering schemes observed in neutron diffraction and NMR spectroscopy. The FM ground state of the In rich alloys, characterized by a FM alignment of Mn$_{\mathrm{4a}}$, Mn$_{\mathrm{4b}}$ and Ni moments, turns for the Sn-rich alloys into a mixed AF/FM ground state, in which the moments of Mn$_{\mathrm{4b}}$ ions are FM or AF coupled to the Mn$_{4a}$ ferromagnetic sublattice depending on whether at least one of the nearest-neighbor 8c sites is occupied by Mn (Fig. \ref{2}a). The chemical statistical disorder leads to the formation of FM and AF clusters on the 4b sites. At the same time, the competition of the various interactions that couple the Mn$_{\mathrm{4b}}$ moments with the Mn$_{4a}$ and $Ni_{8c}$ sublattices reduces the effective inter-sublattices coupling, thus resulting at finite-temperature in the decoupling of the 4b sublattice from the strong FM 4a-8c matrix and a fast relaxation of Mn$_{4b}$-order. By moving towards the pure In-alloy, the FM exchange starts to dominate over the AF interaction, increasing the coupling between the two sublattices and, therefore, the Mn$_{4b}$ onset temperature.

The effective decoupling of the Mn$_{4a}$ and Mn$_{4b}$ sublattice magnetization is confirmed by finite-temperature Monte Carlo simulations of a localized spin model (details are reported in the Supplementary Material \cite{SUP,Muller2019}). The simulations are based on a realistic parameters set obtained from DFT, where only the ratio between the FM ($J_{FM}$ between nearest neighbors Ni and Mn$_{4a,4b}$) and the AF ($J_{AF}$ between next-nearest neighbors Mn$_{4a}$ and Mn$_{4b}$ ) exchange constants was deliberately varied. The good agreement of the simulated results in Fig. \ref{2}b with the experimental phase diagram in Fig. \ref{1}b confirms that the onset temperature of the 4b sublattice depends in first approximation on the $J_{AF}/ J_{FM}$ ratio, which is effectively controlled by the In/Sn substitution. This suggests that the decoupling is not a direct consequence of the magnetic inhomogeneity introduced by the excess Mn on the 8c sites but it is due to a change in the inter-sublattices interactions.

It is now worth discussing the nature of the Mn$_{4b}$ sublattice “transition”. As discussed above, the moments of this sublattice are subjected to strongly frustrated interactions that effectively decouple them from the FM matrix. On reducing the temperature both NMR and neutron diffraction clearly show the development of a long range ordered moment, which the almost linear temperature dependence suggests to be an induced moment. This implies the lack of a thermodynamic transition at the order-temperature, consistent with the symmetry of the system and with the absence of a heat capacity peak in the Monte Carlo simulations (Fig. S19 \cite{SUP}). The ordering of the Mn$_{4b}$ sublattice then resembles the crossover between a paramagnetic state and a fully polarized state in presence of an external magnetic field.

Based on these findings we can derive guidelines to control the magnetic structure of the Ni$_{48}$Mn$_{36}$In$_{16-x}$Sn$_x$  alloys, which is composed of two intertwined, but effectively decoupled, FM sublattices. The Mn$_{4a}$ sublattice is long-range FM ordered and defines the macroscopic Curie temperature of the alloy, which depends mainly on the direct Mn-Ni exchange (and to a lesser extent on the intra-lattice long-range Mn$_{4a}$--Mn$_{4a}$ coupling) and can be amended by excess Mn on the Ni sublattice. In turn, the 4b sublattice, which is only partially occupied by Mn, orders at a temperature, which decreases with  replacing In with Sn. This effective decoupling of magnetic sublattices is due to the superposition of FM (Mn--Ni and Mn$_{4a}$--Mn$_{4a}$) and AF (Mn$_{4a}$--Mn$_{4b}$) interactions and to different magnetic surrounding of Mn$_{4a}$ and Mn$_{4b}$. Indeed, the different number of nearest-neighbors Mn atoms surrounding the 4a and 4b sites unbalances the exchange interactions of Mn$_{4a}$ and Mn$_{4b}$ moments.The strengthening of AF Mn$_{4a}$--Mn$_{4b}$ exchange coupling, obtained by replacing In with Sn, effectively reduces the overall coupling energy of Mn$_{4b}$ moments, thus lowering their onset temperature. In the intermediate temperature range (Fig. \ref{1}b), between the macroscopic Curie temperature and the 4b order-temperature, the Mn$_{4b}$ moments behave like PM spins inside a FM matrix. The Sn-ternary alloy is a borderline configuration, showing a mixed ferro- and ferrimagnetic ground state, experimentally observed only at very low temperatures, caused by the distribution of positive and negative interactions. The strengthening of the AF Mn$_{4a}$--Mn$_{4b}$ exchange coupling can be related to the alterations in the d-metal density of states caused by the variation of the number of $sp$-electrons, as pointed out by our DFT calculations. Nevertheless, the effect of the unit cell reduction, caused by the Sn substitution, cannot be completely excluded. The latter point needs to be further investigated together with the possible role of lattice distortions, that can locally vary the distances between Mn atoms.

In conclusion, combining bulk sensitive experimental results, DFT calculations and Monte Carlo simulations we have shown that it is possible to effectively decouple two FM sublattices thanks to competing interactions in the Ni$_{48}$Mn$_{36}$In$_{16-x}$Sn$_x$ Heusler alloys. The effective ratio of FM and AF exchange couplings, which directly relates to the degree of frustration, is controlled by the electron doping due to the In/Sn substitution. The observed magnetic state is a rare example of magnetic frustration in a highly-symmetric ferromagnetic metallic system. The effective decoupling of the two ferromagnetic sublattices leads to a peculiar magnetic phase at intermediate temperature in which one sublattice is long-range ferromagnetically ordered and the second acts as a paramagnet. Our work shows that magnetic frustration in ferromagnetic metallic systems can lead to interesting and emerging phenomena. Moreover, the deliberate manipulation of competing localized and itinerant exchange interactions through the $p$-electrons of non-spin-polarized main group element adds routes to design new Heusler-type and other functional magnetic materials, increasing their potential applications. 

All data are available from the authors upon reasonable request. Raw data for the neutron diffraction measurements are available at 10.5286/ISIS.E.RB1910102-1.
The authors acknowledge the Science and Technology Facility Council (UK) for the provision of neutron beam time on the WISH diffractometer under the proposal
RB1910102 and Gavin Stenning for help on the PPMS instrument in the Materials Characterization Laboratory at the ISIS Neutron and Muon Source. The large scale DFT calculations were carried out on the Lichtenberg HPC system of the TU-Darmstadt (Project No. 1442 and No. 20039). O.N.M. and M.E.G. gratefully acknowledge financial support from the Deutsche Forschungsgemeinschaft (DFG) within TRR 270 (sub-Project No. B06), Project-ID 405553726. P.B. also acknowledges computing resources provided by CINECA under Project SUPER and ISCRA-B Grant No. IsB20-PRISM, the STFC scientific computing department’s SCARF cluster and the HPC resources at the University of Parma, Italy.


%


\begin{thebibliography}{28}
\bibitem [{Not()}]{Note}%
  \BibitemOpen
  \href@noop {} {\bibinfo {title} {{For clarity in the remainder of the paper
  we will refer to the sublattices using the Wyckoff positions of the
  paramagnetic austenite phase as shown in Fig. \ref{1}a}}}\BibitemShut
  {NoStop}%
\end{thebibliography}

\end{document}